\documentclass[11pt]{article}
\usepackage{a4} 

\usepackage{algorithm}

\usepackage{microtype}
\usepackage{amssymb}
\usepackage{amsmath}
\usepackage{bm}
\usepackage{times}
\usepackage{fancybox}
\usepackage{color}
\usepackage{latexsym}
\usepackage{verbatim}

\newtheorem{theorem}{Theorem}[section]

\newtheorem{lemma}[theorem]{Lemma}

\newtheorem{fact}[theorem]{Fact}

\newtheorem{claim}[theorem]{Claim}

\newtheorem{observation}{Observation}[theorem]
\newtheorem{definition}[theorem]{Definition}

\newtheorem{remark}[theorem]{Remark}

\newcommand{\bproof}{\noindent{\it Proof}}
\newcommand{\cproof}{\noindent{\it Proof of Claim:}}

\newcommand{\eproof}{\hspace*{\fill}$\rule{2mm}{2mm}$~~~~~\bigskip}

\renewenvironment{proof}{\bproof. }{\eproof}

\newenvironment{claimproof}{\cproof. }{\hspace*{\fill}\vspace{5mm}}
\usepackage{thm-restate}
\usepackage{hyperref}

\usepackage{cleveref}

\newcommand{\op}[1]{\ensuremath{\operatorname{#1}}}

\newcommand{\mini}{\op{mini}}

\newcommand{\CNF}{\mbox{\small\rm 3-CNF}}

\newcommand{\XP}{\mbox{\small\rm XP}}
\newcommand{\FPT}{\mbox{\small\rm FPT}}
\newcommand{\MINI}{\mbox{\small\rm MINI{[}1{]}}}
\newcommand{\msat}{\mbox{\small\rm MINI{-}3SAT}}
\newcommand{\motsat}{\mbox{\small\rm MINI-1-in-3 POSITIVE 3-SAT}}
\newcommand{\kLINEQ}{\mbox{\small\rm \textit{k}{-}\textsc{Lin{-}Eq}}}
\newcommand{\otsat}{\mbox{\small\rm 1-in-3 POSITIVE 3-SAT}}
\newcommand{\REM}{\mbox{\small\rm REM}}

\newcommand{\NP}{\mbox{\rm NP}}

\newcommand{\coNP}{\mbox{\rm coNP}}

\newcommand{\AM}{\mbox{\rm AM}}

\newcommand{\TRUE}{\mbox{\rm TRUE}}
\newcommand{\FALSE}{\mbox{\rm FALSE}}

\newcommand{\EXPSPACE}{\mbox{\rm EXPSPACE}}

\newcommand{\HN}{\mbox{\rm HN}}

\newcommand{\cequalp}{\ensuremath{\rm C_{=}P}}
\newcommand{\poly}{\mbox{\rm\small poly}}

\newcommand{\Q}{\mathbb{Q}}
\newcommand{\Z}{\mathbb{Z}}

\newcommand{\F}{\ensuremath{\mathbb{F}}}

\renewcommand{\v}{\mbox{\bf v}}

\renewcommand{\angle}[1]{\langle #1\rangle}

\newcommand{\X}{\mbox{\rm X}}

\newcommand{\C}{\mathbb{C}}

\newcommand{\W}{\mbox{\rm W}}

\newcommand{\perm}{\mbox{\textrm Perm}}

\newcommand{\Divide}{\mbox{\textrm Divide}}

\renewcommand{\angle}[1]{\langle #1\rangle}

  \title{Univariate Ideal Membership Parameterized by Rank, Degree, and Number of Generators}  

\author{V. Arvind\thanks{Institute of Mathematical Sciences (HBNI), Chennai,
India, \texttt{email: arvind@imsc.res.in}}  \and Abhranil Chatterjee\thanks{Institute of Mathematical Sciences (HBNI), Chennai,
    India, \texttt{email: abhranilc@imsc.res.in}} \and Rajit Datta\thanks{Chennai Mathematical Institute, Chennai, India, \texttt{email: rajit@cmi.ac.in}} \and Partha
  Mukhopadhyay\thanks{Chennai Mathematical Institute, Chennai, India,
    \texttt{email: partham@cmi.ac.in}}
}

\begin{document}

\maketitle

\begin{abstract}
 Let $\F[\X]$ be the polynomial ring over the variables $\X=\{x_1,x_2,
 \ldots, x_n\}$. An ideal $I=\angle{p_1(x_1), \ldots, p_n(x_n)}$
 generated by univariate polynomials $\{p_i(x_i)\}_{i=1}^n$ is a
 \emph{univariate ideal}. We study the ideal membership problem for
 the univariate ideals and show the following results.
 
 \begin{itemize}
 \item Let $f(\X)\in\F[\ell_1, \ldots, \ell_r]$ be a (low rank)
   polynomial given by an arithmetic circuit where $\ell_i : 1\leq
   i\leq r$ are linear forms, and $I=\angle{p_1(x_1), \ldots,
     p_n(x_n)}$ be a univariate ideal. Given $\vec{\alpha}\in \F^n$,
   the (unique) remainder $f(\X) \pmod I$ can be evaluated at
   $\vec{\alpha}$ in deterministic time $d^{O(r)}\cdot \poly(n)$,
   where $d=\max\{\deg(f),\deg(p_1)\ldots,\deg(p_n)\}$. This yields an
   $n^{O(r)}$ algorithm for minimum vertex cover in graphs with
   rank-$r$ adjacency matrices. It also yields an $n^{O(r)}$ algorithm
   for evaluating the permanent of a $n\times n$ matrix of rank $r$,
   over any field $\F$. Over $\Q$, an algorithm of similar run time
   for low rank permanent is due to Barvinok~\cite{Bar96} via a
   different technique.

\item Let $f(\X)\in\F[\X]$ be given by an arithmetic circuit of degree
  $k$ ($k$ treated as fixed parameter) and $I=\angle{p_1(x_1), \ldots,
  p_n(x_n)}$. We show that 
  in the special case when $I=\angle{x_1^{e_1}, \ldots,
    x_n^{e_n}}$, we obtain a randomized $O^*(4.08^k)$ algorithm that
  uses $\poly(n,k)$ space.

\item Given $f(\X)\in\F[\X]$ by an arithmetic circuit and
  $I=\angle{p_1(x_1), \ldots, p_k(x_k)}$, membership testing is
  $\W[1]$-hard, parameterized by $k$. The problem is $\MINI$-hard in
  the special case when $I=\angle{x_1^{e_1}, \ldots, x_k^{e_k}}$.


 \end{itemize}  
  \end{abstract}

 \section{Introduction}\label{intro}
 
 Let $R=\F[x_1, x_2, \ldots, x_n]$ \footnote{We often use the
   shorthand notation $\F[\X]$. } be the ring of polynomials over the
 variables $\X=\{x_1, x_2, \ldots, x_n\}$. A subring $I\subseteq R$ is
 an ideal if $ I R\subseteq I$. Computationally, an ideal $I$ is often
 given by generators : $I = \angle{f_1,f_2,\ldots,f_{\ell}}$. Given
 $f\in R$ and $I=\angle{f_1, \ldots, f_{\ell}}$, the \emph{Ideal
   Membership problem} is to decide whether $f\in I$ or not. In
 general, this is computationally highly intractable. In fact, it is
 $\EXPSPACE$-complete even if $f$ and the generators $f_i,i\in[\ell]$
 are given explicitly by sum of monomials \cite{MM82}. Nevertheless,
 special cases of ideal membership problem have played important roles
 in several results in arithmetic complexity. For example, the
 polynomial identity testing algorithm for depth three
 $\Sigma\Pi\Sigma$ circuits with bounded top fan-in; the structure
 theorem for $\Sigma\Pi\Sigma(k,d)$ identities use ideal membership
 very crucially \cite{AM10, KS07, SS13}.

In this paper, our study of ideal membership is motivated by a basic
result in algebraic complexity : the Combinatorial Nullstellensatz of
Alon \cite{alon99}, and we recall a basic result in that paper.

\begin{theorem}\label{alon}
Let $\F$ be any field, and $f(\X)\in\F[\X]$. Define polynomials
$g_i(x_i) = \prod_{s\in S_i}(x_i - s)$ for nonempty subsets $S_i, 1\le
i\le n$ of $\F$. If $f$ vanishes on all the common zeros of $g_1,
\ldots, g_n$, then there are polynomials $h_1, \ldots, h_n$ satisfying
$\deg(h_i)\leq \deg(f) - \deg(g_i)$ such that $f=\sum_{i=1}^n h_ig_i$.
\end{theorem}

The theorem can be restated in terms of ideal membership: Let $f(\X)
\in\F[\X] $ be a given polynomial, and
$I=\angle{g_1(x_1),g_2(x_2),\ldots,g_n(x_n)}$ be an ideal generated by
univariate polynomials $g_i$ \emph{without repeated roots}. Let
$Z(g_i)$ denote the zero set of $g_i, 1\leq i \leq n$. By
Theorem~\ref{alon}, if $f\not\in I$ then there is a
$\vec{\alpha}=(\alpha_1, \ldots, \alpha_n)\in Z(g_1)\times \cdots
\times Z(g_n)$ such that $f(\vec{\alpha})\neq 0$. Of course, if $f\in
I$ then $f|_{Z(g_1)\times\cdots\times Z(g_n)}=0$. 

Ideals $I$ generated by univariate polynomials are called
\emph{univariate ideals}. For any univariate ideal $I$ and any
polynomial $f$, by repeated application of the division algorithm, we
can write $f(\X)=\sum_{i=1}^n h_i(\X) g_i(x_i) + R(\X)$ where $R$ is
unique and for each $i\in [n] : \deg_{x_i}(R) < \deg(g_i(x_i))$. Since
the remainder is unique, it is convenient to write $R = f \mod I$. By
Alon's theorem, if $f\not\in I$ then there is a $\vec{\alpha}\in
Z(g_1) \times \cdots \times Z(g_n)$ such that $R(\vec{\alpha})\neq 0$.

As an application of the theorem, Alon and Tarsi showed that checking
$k$-colorability of a graph $G$ is polynomial-time equivalent to
testing whether the graph polynomial $f_{G}$ is in the ideal
$\angle{x_1^{k}-1, \ldots, x_n^{k}-1}$ \cite{alon99}. It follows that
univariate ideal membership problem $\coNP$-hard.

Univariate ideal membership is further motivated by its connection
with two well-studied problems. Computing the permanent of a $n\times
n$ matrix over any field $\F$ can be cast in terms of univariate ideal
membership. Given a matrix $A=(a_{i,j})_{1\leq i,j\leq
  n}\in\F^{n\times n}$, consider the product of linear forms $P_A(\X)
= \prod_{i=1}^{n} (\sum_{j=1}^n a_{ij} x_j)$. The following
observation is well known.

\begin{fact}\label{perm}
The permanent of the matrix $A$ is given by the coefficient of the
monomial $x_1x_2\cdots x_n$ in $P_A$.
\end{fact}   

It follows immediately that $P_A(\X) \pmod{\angle{x_1^2, \ldots, x_n^2}}= 
\perm(A) ~x_1 x_2 \cdots x_n$.  I.e., the remainder $P_A
\pmod{\angle{x_1^2, \ldots, x_n^2}}$ evaluates to $\perm(A)$ at the
point $\vec{1}\in\F^n$.

Next, we briefly mention the connection of univariate ideal membership
with the multilinear monomial detection problem, a benchmark problem
that is useful in designing fast parameterized algorithms for a host
of problems~\cite{Kou08, Kou12,KW16,RW09}.

Notice that, given an arithmetic circuit $C$ computing a polynomial
$f\in\F[\X]$ of degree $k$, checking if $f$ has a nonzero multilinear
monomial of degree $k$ is equivalent to checking if $f
\pmod{\angle{x_1^2, \ldots, x_n^2}}$ is nonzero. Moreover, the
constrained multilinear detection problem studied in~\cite{BKK16,
  Kou12} can also be viewed as a problem of deciding membership in a
univariate ideal.

\vspace{-.25cm}
\subsection{Our Results}
	
A contribution of this paper is to consider several parameterized
problems in arithmetic complexity as instances of univariate ideal
membership.  One parameter of interest is the rank of a multivariate
polynomial: We say $f\in\F[\X]$ is a \emph{rank $r$} polynomial if $f
\in \F[\ell_1 , \ell_2 ,\ldots,\ell_r]$ for linear forms $\ell_j :
1\leq j\leq r$. This concept has found application in algorithms for
depth-3 polynomial identity testing \cite{SS13}. Given a univariate
ideal $I$, a point $\vec{\alpha}\in\F^n$, and an arithmetic circuit
computing a polynomial $f$ of rank $r$, we obtain an efficient
algorithm to compute $f\pmod{I}$ at $\vec{\alpha}$.
 
 \begin{theorem}\label{main-thm-4}
Let $\F$ be an arbitrary field where the field arithmetic can be done
efficiently, and $C$ be a polynomial-size arithmetic circuit computing
a polynomial $f$ in $\F[\ell_1,\ell_2,\ldots,\ell_r]$, where $\ell_1,
\ell_2, \ldots, \ell_r$ are given linear forms in $\{x_1,x_2,\ldots,x_n\}$.
Let $I=\angle{p_1, \dots, p_n}$ be a univariate ideal generated by
$p_i(x_i)\in\F[x_i], 1\le i\le n$. Given $\vec{\alpha}\in \F^n$, we
can evaluate the remainder $f \pmod{I}$ at the point $\vec{\alpha}$ in
time $d^{O(r)} \poly(n)$, where $d=\max\{\deg(f),\deg(p_i): 1\le i\le
n\}$.
\end{theorem}

This also allows us to check whether $f\in I$ by picking a point
$\vec{\alpha}$ at random and checking whether $f\pmod{I}$ evaluated at
$\vec{\alpha}$ is zero or not. The intuitive idea behind the proof of
Theorem \ref{main-thm-4} is as follows.

Given a polynomial $f(\X) \in \F[\ell_1, \ldots, \ell_r]$, a
univariate ideal $I=\angle{p_1(x_1), \ldots, p_n(x_n)}$, and a point
$\vec{\alpha}\in\F^n$, we first find an invertible linear
transformation $T$ such that the polynomial $T(f)$ becomes a
polynomial over at most $2r$ variables. Additionally $T$ has the
property that $T$ fixes the variables $x_1, \ldots, x_r$. Then we
recover the polynomial (call it $\tilde{f}$) over at most $2r$
variables explicitly and perform division algorithm with respect to
the ideal $I_{[r]}=\angle{p_1(x_1), \ldots, p_r(x_r)}$. For notational convenience, call
$\tilde{f}$ be the polynomial obtained over at most $2r$ variables.
It turns out $T^{-1}(\tilde{f})$ is the \emph{true remainder} $f
\pmod{I_{[r]}}$. Since the variables $x_1, \ldots, x_r$ do not play
role in the subsequent stages of division, we can eliminate them by
substituting $x_i \leftarrow \alpha_i$ for each $1\leq i\leq r$. Then
we apply the division algorithm on $T^{-1}(\tilde{f})|_{x_i \leftarrow
  \alpha_i : 1\leq i\leq r}$ recursively with respect to the ideal
$I_{[n]\setminus[r]}$ to compute the final remainder at the point
$\vec\alpha$.

Our next result is an efficient algorithm to detect vertex cover in
low rank graphs.  A graph $G$ is said to be of rank $r$ if the rank of
the adjacency matrix $A_G$ is of rank $r$. Graphs of low rank were
studied by Lovasz and Kotlov~\cite{KL96,Kot97} in the context of graph
coloring. Our idea is to construct a low rank polynomial from the
graph and check its membership in an appropriate univariate ideal.

\begin{theorem}\label{lowVC}
 Given a graph $G=(V,E)$ on $n$ vertices such that the rank of the
 adjacency matrix $A_{G}$ is at most $r$, and a parameter $k$, there
 is a randomized $n^{O(r)}$ algorithm to decide if the graph $G$
 has vertex cover of size $k$ or not.
\end{theorem}
 
Theorem~\ref{main-thm-4} also yields an $n^{O(r)}$ algorithm to
compute the permanent of rank-$r$ matrices over any field. Barvinok
had given~\cite{Bar96} an algorithm of same running time for the
permanent of low rank matrices (over $\Q$) using apolar bilinear
forms. By Fact~\ref{perm}, if matrix $A$ is rank $r$ then $P_A$ is a
rank-$r$ polynomial, and for the univariate ideal $I=\angle{x_1^2,
  \ldots, x_n^2}$ computing $P_A \pmod{I}$ at the point ${\vec{1}}$
yields the permanent. Theorem~\ref{main-thm-4} works more generally
for all univariate ideals. In particular, the ideal in the proof of
Theorem \ref{lowVC} is generated by polynomials that are not powers of
variables. Thus, Theorem~\ref{main-thm-4} can potentially have more
algorithmic consequences than the technique in \cite{Bar96}.

If $k$ is the degree of the
input polynomial and the ideal is given by the powers of variables as
generators, we have a randomized $\FPT$ algorithm for the problem.

 \begin{theorem}\label{degparalgo}
 Given an arithmetic circuit $C$ computing a polynomial $f(\X) \in
 \mathbb{Z}[\X]$ of degree $k$ and integers $e_1,e_2,\ldots,e_n$,
 there is a randomized algorithm to decide whether $f \not\in
 \angle{x^{e_1}_1,x^{e_2}_2,\ldots,x^{e_n}_n}$ in $O^*(4.08^k)$
 time.
\end{theorem}

Note that this generalizes the well-known problem of \emph{multilinear
  monomial detection} for which the ideal of interest would be
$I=\angle{x^2_1,x^2_2,\ldots,x^2_n}$. Surprisingly, the run time of
the algorithm in Theorem~\ref{degparalgo} is independent of the
$e_i$. Brand et al.\ have given the first FPT algorithm for
multilinear monomial detection in the case of general circuit with run
time randomized $O^*(4.32^k)$~\cite{BDH18}. Recently, this problem has
also been studied using the Hadamard product~\cite{ACDM18} of the
given polynomial with the elementary symmetric polynomial (and
differently using apolar bilinear forms~\cite{Pra18}). Our proof of
Theorem~\ref{degparalgo} shows that checking membership of $f$ in the
ideal $\angle{x^{e_1}_1,\ldots,x^{e_n}_n}$ is efficiently reducible to
computing the (scaled) Hadamard product of $f$ with a modified
elementary symmetric polynomial.

When the number of generators in the ideal is treated as the fixed
parameter, the problem is $\W[1]$-hard.
 
 \begin{theorem}\label{genparhard}
 Given a polynomial $f(\X) \in \mathbb{F}[\X]$ by an arithmetic
 circuit $C$ and univariate polynomials
 $p_1(x_1),p_2(x_2),\ldots,p_k(x_k)$, checking if $f \not \in
 \angle{p_1(x_1),p_2(x_2),\dots,p_k(x_k)}$ is $\W[1]$-hard with $k$ as
 the parameter.
\end{theorem}

Theorem~\ref{genparhard} is shown by a suitable reduction from
independent set problem to ideal membership. To find an independent
set of size $k$, the reduction produces an ideal with $k$ univariates
and the polynomial created from the graph has $k$ variables. Unlike
Theorem \ref{degparalgo}, the above parameterization of the problem
remains $\MINI$-hard even if the ideal is generated by powers of
variables. More precisely, we show the following result.

\begin{theorem}\label{genparuna}
Let $C$ be a polynomial-size arithmetic circuit computing a polynomial
$f\in\F[\X]$. Let $I =
\angle{{x_1}^{e_1},{x_2}^{e_2},\ldots,{x_k}^{e_k}}$ be the given ideal
where $e_1, \ldots, e_k$ are given in unary. Checking if $f\not\in I$
is $\MINI$-hard with $k$ as parameter.
\end{theorem}

It turns out that the complement of the ideal membership problem can
be easily reduced from $\kLINEQ$ problem which asks if there is a
$\vec{x}\in\{0,1\}^n$ satisfying $A\vec{x} = \vec{b}$, where
$A\in\F^{k\times n}$ and $\vec{b}\in\F^k$.

We can show $\kLINEQ$ is hard for the parameterized complexity class
$\MINI$ by reducing the miniature version of $\otsat$ to it.

As already mentioned, the result of Alon and Tarsi \cite{alon99} shows
that the membership of $f_G$ in $\angle{x_1^k-1, \ldots, x_n^k-1}$ is
$\coNP$-hard and the proof crucially uses the fact that the roots of
the generator polynomials are all distinct.  This naturally raises the
question if univariate ideal membership is in $\coNP$ when each
generator polynomial has distinct roots. We show membership in
$\coNP$.
  
 \begin{theorem}\label{main-thm-1}
 Let $f\in\Q[\X]$ be a polynomial of degree at most $d$ given by a
 black-box. Let $I=\angle{p_1(x_1), \ldots, p_n(x_n)}$ be an ideal
 given explicitly by a set of univariate polynomials $p_1, p_2,
 \ldots, p_n$ as generators of maximum degree bounded by $d$. Let $L$
 be the bit-size upper bound for any coefficient in $f, p_1, p_2,
 \ldots, p_n$. Moreover, assume that $p_i$s have distinct roots over
 $\C$. Then there is a non-deterministic algorithm running in time
 $\poly(n, d, L)$ that decides the non-membership of $f $ in the ideal
 $I$.
 \end{theorem}

\begin{remark}
The distinct roots case discussed in Theorem~\ref{main-thm-1} is in
stark contrast to the complexity of testing membership of $P_A(\X)$ in
the ideal $\angle{x_1^2, \ldots, x_n^2}$. That problem is equivalent
to checking if $\perm(A)$ is nonzero for a rational matrix $A$, which
is hard for the exact counting class C$_{=}$P. Hence it cannot be in
$\coNP$ unless the polynomial-time hierarchy collapses.
\end{remark}
 
 Recall from Alon's Nullstellensatz that if $f\not\in I$, then there
 is always a point $\vec{\alpha}\in Z(p_1)\times\ldots\times Z(p_n)$
 such that $f(\vec\alpha)\neq 0$. Notice that in general the roots
 $\alpha_i \in \C$ and in the standard \emph{Turing Machine} model the
 $\NP$ machine can not guess the roots directly with only finite
 precision. But we are able to prove that the $\NP$ machine can guess
 the tuple of roots $\vec{\tilde{\alpha}}\in\Q^n$ using only
 polynomial bits of precision and still can decide the
 non-membership. The main technical idea is to compute efficiently a
 parameter $M$ only from the input parameters such that
 $|f(\vec{\tilde{\alpha}})|\leq M$ if $f\in I$, and
 $|f(\vec{\tilde{\alpha}})|\geq 2M$ if $f\not\in I$. The $\NP$ machine
 decides the non-membership according to the final value of
 $|f(\vec{\tilde{\alpha}})|$.  We remark that Koiran has considered
 the weak version of Hilbert Nullstellensatz ($\HN$) problem
 \cite{koi96}. The input is a set of multivariate polynomials $f_1,
 f_2, \ldots, f_m \in \Z[\X]$ and the problem is to decide whether
 $1\in\angle{f_1, \ldots, f_m}$. The result of Koiran shows that
 $\overline{\HN}\in \AM$ (under GRH), and it is an outstanding open
 problem problem to decide whether $\overline{\HN}\in \NP$.

 \vspace{-0.25 cm}
\subsubsection*{Organization} 
In Section \ref{prelim} we give some background results. We prove
Theorem \ref{main-thm-4} and Theorem \ref{lowVC} in Section
\ref{barvinokgeneral}.

In Section~\ref{parameterized-section}, we explore the parameterized
complexity of univariate ideal membership. In the first subsection, we
prove 
\ref{degparalgo}, and in the
second subsection we prove Theorems~\ref{genparhard} and
\ref{genparuna}. Finally, in Section~\ref{main-thm1-sec}, we prove
Theorem \ref{main-thm-1}. Several proofs are given in the appendix.

\section{Preliminaries}\label{prelim}
 
\subsubsection*{Basics of Ideal Membership}\vspace{-0.15cm}

Let $\F[\X]$ be the ring of polynomials $\F[x_1,x_2,\ldots,x_n]$. Let
$I\subseteq \F[\X]$ be an ideal given by a set of generators
$I=\angle{g_1, \ldots, g_{\ell}}$. Then for any polynomial
$f\in\F[\X]$, it is a member of the ideal if and only if
$f=\sum_{i=1}^{\ell} h_i g_i$ where $\forall i :
h_i\in\F[\X]$. Dividing $f$ by the $g_i$ by applying the standard
division algorithm does not work in general to check if $f\in
I$. Indeed, the remainder is not even uniquely defined. However, if
the leading monomials of the generators are already pairwise
relatively prime, then we can apply the division algorithm to compute
the unique remainder.
\begin{theorem}[See\cite{CLO07}, Theorem 3, proposition 4, pp.101]
\label{GB-Syz} 
Let $I$ be a polynomial ideal given by a basis $G = \{g_1, g_2,
\cdots, g_s\}$ such that all pairs $i\neq j$ $LM(g_i)$ and $LM(g_j)$
are relatively prime. Then $G$ is a Gr\"obner basis for $I$.
\end{theorem}

In particular, if the ideal $I$ is a univariate ideal given by
$I=\angle{p_1(x_1), \ldots, p_n(x_n)}$, we can apply the division
algorithm to compute the unique remainder $f \pmod{I}$. To bound the
run time of this procedure we note the following: Let $\bar{p}$ denote
the ordered list $\{p_1,p_2,\ldots,p_n\}$. Let $\Divide(f ; \bar{p})$
be the procedure that divides $f$ by $p_1$ to obtain remainder $f_1$,
then divides $f_1$ by $p_2$ to obtain remainder $f_2$, and so on to
obtain the final remainder $f_n$ after dividing by $p_n$. We note the
following time bound for $\Divide(f ; \bar{p})$.

\begin{fact}[See \cite{Sudan98}, Section 6, pp.5-12]\label{div-run-time}
Let $f\in\F[\X]$ be given by a size $s$ arithmetic circuit and
$p_i(x_i)\in\F[x_i]$ be given univariate polynomials. The running time
of $\Divide(f ; \bar{p})$ is bounded by $O(s\cdot \prod_{i=1}^n (d_i +
1)^{O(1)})$, where $d_i=\max\{\deg_{x_i}(f),\deg(p_i(x_i))\}$.
\end{fact} 

\vspace{-0.35cm}
\subsubsection*{On Roots of Univariate Polynomials}\vspace{-0.15cm}

The following lemma shows that the absolute value of any root of a univariate polynomial can be bounded in terms of the degree and the coefficients. The result is folklore.  

\begin{lemma}\label{lemma-1}
 Let $f(x) = \sum_{i=0}^d a_i x^i\in\Q[x]$ be a univariate polynomial and $\alpha$ be a root of $f$. Then, either $\frac{|a_0|}{\sum_{i=1}^d |a_i|}\leq |\alpha|<1$ or $1\leq |\alpha|\leq d \cdot \frac{\max_i |a_i|}{|a_d|}$. 
  \end{lemma}
  
  \begin{proof}
  Since $\alpha$ is a root of $f$, we have that,  $0=f(\alpha)=\sum_{i=0}^d a_i  \alpha^i=0$, and 
  $\sum_{i=1}^d a_i \alpha^i = -a_0$. Then by an application of triangle inequality, we get that 
  $\sum_{i=1}^d |a_i| |\alpha|^i \geq |a_0|$. Now we analyse two different cases. 
 In the first case assume that $|\alpha| < 1$.  
 Observe that $|\alpha| \cdot (\sum_{i=1}^d |a_i|) \geq |a_0|$, and hence $|\alpha|\geq \frac{|a_0|}{\sum_{i=1}^d |a_i|}$. 
  In the second case $|\alpha|\geq 1$. 
   Observe that  $-a_d \alpha^d = \sum_{i=0}^{d-1} a_i \alpha^i$. Then use triangle inequality to get that 
 $|a_d| |\alpha|^d \leq |\alpha|^{d-1} \cdot (\sum_{i=0}^{d-1} |a_i|)$. Now we get the following, 
  $
  |\alpha| \leq \frac{\sum_{i=0}^{d-1} |a_i|} {|a_d|}\leq d \cdot \frac{\max_i |a_i|}{|a_d|}. 
  $
 The lemma follows by combining the two cases.  
   \end{proof}
  
  The next lemma shows that the separation between two distinct roots of any univariate polynomial can be lower bounded in terms of degree and the size of the coefficients. This was shown by Mahler \cite{Mah64}.  
  
  \begin{lemma}\label{lemma-2}
  Let $g(x) = \sum_{i=0}^d a_i x^i \in \Q[x$ and $2^{-L}\leq |a_i|\leq 2^L$ (if $a_i \neq 0$). Let $\alpha, \beta$ are two distinct roots of $g$. Then $|\alpha-\beta| \geq \frac{1}{2^{O(d L)}}$. 
  \end{lemma} 

The following lemma states that any univariate polynomial can not get a very small value (in absolute sense) on any point which is far from every root. 

 \begin{lemma}\label{npguess}
  Let $f= \sum_{i=1}^d a_i x^i$ be a univariate polynomial with $2^{-L} \leq |a_i| \leq 2^{L}$ (if $a_i \neq 0$).
  Let $\tilde{\alpha}$ be a point such that $|\tilde{\alpha} - \beta_i| \geq \delta$ for every root $\beta_i$ of $f$ then $|f(\tilde{\alpha})| \geq 2^{-L} \delta^d$.
  
 \end{lemma}

\begin{proof}
 We observe that,
$f(\tilde{\alpha}) = c \prod_{i=1}^{d} (\tilde{\alpha} - \beta_i)$.
Since $|\tilde{\alpha} - \beta_i| \geq \delta$ we get,
$|f(\tilde{\alpha})| = |c| \prod_{i=1}^{d} |\tilde{\alpha} - \beta_i| \geq 2^{-L} \delta^d$. 
 This completes the proof.
\end{proof} 
\vspace{-0.3cm}
\subsubsection*{Parameterized Complexity Classes}\vspace{-0.15cm}

We recall some standard definitions in parameterized Complexity
\cite[ch.1,pp. 7-14]{FKLMPPS15}. We only state them informally. For a
parameterized input problem $(x,k)$ with $k$ be the parameter of
interest, we say that the problem is in $\FPT$ if it has an algorithm
with run time $f(k) |(x,k)|^{O(1)}$ for some computable function $f$.
A parameterized reduction \cite[def. 13.1]{FKLMPPS15} between two
problems should be computable in time $f(k) |(x,k)|^{O(1)}$, and if
the reduction outputs $(x',k')$ then $k' \leq f(k)$. A parameterized
problem is in the class $\XP$ if it has an algorithm with run time
$|x|^{f(k)}$ for some computable function $f$.

For the purpose of this paper, it suffices to note that a
parameterized problem $L$ is in the class $\W[1]$ if there is a
parameterized reduction from $L$ to some standard $\W[1]$-complete
problem like, e.g., the $k$-Independent set problem (more details
can be found in, e.g, \cite[def. 13.16]{FKLMPPS15}). 

The complexity class $\MINI$ consists of parameterized problems that
are miniature versions of $\NP$ problems: For $L\in\NP$, its miniature
version $\mini(L)$ has instances of the form $(0^n,x)$, where $|x|\le
k\log n$, $k$ is the fixed parameter, and $x$ is an instance of
$L$. Showing $\mini(L)$ to be $\MINI$-hard under parameterized
reductions is evidence of its parameterized intractability, for it
cannot be in $\FPT$ assuming the Exponential Time
Hypothesis~\cite{DEFPR03}.

\section{Ideal Membership for Low Rank Polynomials}\label{barvinokgeneral}

In this section we prove Theorem \ref{main-thm-4}. Given a $r$-rank
polynomial $f$ by an arithmetic circuit, a univariate ideal $I$, and a
point $\vec{\alpha}\in\F^n$, we give an $n^{O(r)}$ time algorithm to
evaluate the remainder polynomial $f\pmod I$ at $\vec{\alpha}$. As
mentioned in Section \ref{intro}, an application of our result yields
an $n^{O(r)}$ time algorithm for computing the permanent of rank-$r$
matrices over any field. Barvinok \cite{Bar96}, via a different
method, had obtained an $n^{O(r)}$ time algorithm for this problem
over $\Q$. We also obtain an $n^{O(r)}$ time algorithm for minimum
vertex cover of low rank graphs. We first define the notion
\emph{rank} of a polynomial in $\F[\X]$.

\begin{definition}\label{low-rank}
A polynomial $f(\X)\in\F[\X]$ is a \emph{rank-$r$ polynomial} if there
are linear forms $\ell_1, \ell_2, \ldots, \ell_r$ such that $f(\X)$ is
in the sub-algebra $\F[\ell_1, \ldots, \ell_r]$.
\end{definition} 

For an unspecified fixed parameter $r$, we refer to rank-$r$
polynomials as \emph{low rank polynomials}.

Given $\vec{\alpha}\in\F^n$, a univariate ideal $I=\angle{p_1(x_1),
  \ldots, p_n(x_n)}$, and a rank $r$ polynomial $f(\ell_1, \ldots,
\ell_r)$ we show how to compute $f(\ell_1, \ldots, \ell_r) \pmod I$ at
$\vec\alpha$ using a recursive procedure $\REM(f(\ell_1, \ldots,
\ell_r), I, \vec\alpha)$ efficiently.  We introduce the following
notation. For $S\subseteq [n]$, the ideal $I_S =\angle{p_i(x_i) : i\in
  [S]}$.

We first observe the following lemma which shows how to remove the
redundant variables from a low rank polynomial.
 
 \begin{lemma}\label{varsep}
  Given a polynomial $f(\ell_1,\ldots,\ell_r)$ where
  $\ell_1,\ldots,\ell_r$ are linear forms in $\F[\X]$, there is an
  invertible linear transform $T: \mathbb{F}^{n} \mapsto
  \mathbb{F}^{n}$ that fixes $x_1, \ldots, x_r$ and the transformed
  polynomial $T(f)$ is over at most $2r$ variables.
  \end{lemma}

  \begin{proof}
   Write each linear form $\ell_i$ in two parts: $\ell_i = \ell_{i,1}
   + \ell_{i,2}$, where $\ell_{i,1}$ is the part over variables
   $x_1,\ldots,x_r$ and $\ell_{i,2}$ is over variables
   $x_{r+1},\ldots,x_n$. W.l.o.g, assume that
   $\{\ell_{i,2}\}^{r'}_{i=1}$ is a maximum linearly independent
   subset of linear forms in $\{\ell_{i,2}\}^{r}_{i=1}$. Let
   $T:\F^n\rightarrow\F^n$ be the invertible linear map that fixes
   $x_1, \ldots, x_r$, maps the independent linear forms $\{\ell_{i,2}
   \}^{r'}_{i=1}$ to variables $x_{r+1},\ldots,x_{r+r'}$, and suitably
   extends $T$ to an invertible map. This completes the proof.
\end{proof}

The following lemma shows that the univariate division and evaluating
the remainder at the end can be achieved by division and evaluation
partially.

\begin{lemma} \label{partial}
Let $f(\X)\in\F[\X]$ and $I = \angle{p_1(x_1), \ldots, p_n(x_n)}$ be a
univariate ideal. Let $R(\X)$ be the unique remainder $f \pmod I$. Let
$\vec{\alpha}\in\F^r, r\leq n$ and $R_r(\X) = f \pmod{I_{[r]}}$. Then
$R(\alpha_1, \ldots, \alpha_r, x_{r+1}, \ldots, x_n) = R_r(\alpha_1,
\ldots, \alpha_r, x_{r+1}, \ldots, x_n) \pmod{I_{[n]\setminus[r]}}$.
\end{lemma}

We require the following lemma in the proof of the main result of this
section.

\begin{lemma} \label{inv}
Let $f\in\F[\X]$, and $T : \F^n\rightarrow \F^n$ be an invertible
linear transformation fixing $x_1, \ldots, x_r$ and mapping $x_{r+1},
\ldots, x_n$ to linearly independent linear forms over $x_{r+1},
\ldots, x_n$.  Write $R = f \pmod{I_{[r]}}$ and $R' = T(f)
\pmod{I_{[r]}}$. Then $R' = T(R)$.
\end{lemma}
 
The proofs of Lemmas \ref{partial} and \ref{inv} are given 
in Section~\ref{app1} of the appendix.

\subsubsection{Proof of Theorem~\ref{main-thm-4}}

\begin{toneproof}
We now describe a recursive procedure $\REM$ to solve the problem. The
initial call to it is $\REM(f(\ell_1, \ldots, \ell_r),
I_{[n]},\vec\alpha)$. We apply the invertible linear transformation
obtained in Lemma \ref{varsep} to get the polynomial $T(f)$ over the
variables $x_1, \ldots, x_r, x_{r+1}, \ldots, x_{r+r'}$ where $r'\leq
r$.\footnote{We use $f$ to denote $f(\ell_1, \ldots, \ell_r)$.} The
polynomial $T(f)$ can be explicitly computed in time
$\poly(L,s,n,d^{O(r)})$. Then we compute the remainder polynomial
$f'(x_1, \ldots, x_{r + r'}) = T(f) \pmod{I_{[r]}}$ by applying the
division algorithm which runs in time $\poly(L,s,n,d^{O(r)})$.  Next
we compute the polynomial $g=f'(\alpha_1, \ldots, \alpha_r, x_{r+1},
\ldots, x_{r+r'})$.  Notice from Lemma \ref{varsep} that $T^{-1}(x_{r
  + i}) = \ell_{i,2}$ for $1\leq i\leq r'$, thus we are interested in
the polynomial $g(\ell_{1,2}, \ldots, \ell_{r',2})$. Now we
recursively compute $\REM(g(\ell_{1,2}, \ldots, \ell_{r',2}),
I_{[n]\setminus[r]},\vec\alpha')$ where $\vec\alpha' =
(\alpha_{r+1},\ldots,\alpha_n)$.

\subsubsection*{Correctness of the algorithm.}   

Let $R(\X) = f \pmod{I_{[n]}}$ be the unique remainder polynomial. Let
$R_r(\X) = f\pmod{I_{[r]}}$ and we know that $R_r
\pmod{I_{[n]\setminus [r]}} = R$. So by Lemma \ref{partial}, to show
the correctness of the algorithm, it is enough to show that
$g(\ell_{1,2}, \ldots, \ell_{r',2}) = R_r(\alpha_1, \ldots, \alpha_r,
x_{r+1}, \ldots, x_n)$.

Following Lemma \ref{inv}, write $R'=f'(x_1,\ldots, x_r, x_{r+1},
\ldots, x_n) = T(f)\pmod{I_{[r]}}$.  Then, by Lemma \ref{inv} we
conclude that $R' = T(R_r)$. It immediately follows that $R_r=
T^{-1}(R')=f'(x_1,\ldots, x_r, T^{-1}(x_{r+1}), \ldots,
T^{-1}(x_n))$. Now by definition the polynomial $g(\ell_{1,2}, \ldots,
\ell_{r',2})$ is $f'(\alpha_1, \ldots, \alpha_r, T^{-1}(x_{r+1}),
\ldots, T^{-1}(x_{r+r'}))$ which is simply $R_r(\alpha_1, \ldots,
\alpha_r, x_{r+1}, \ldots, x_n)$.
\vspace{-0.35 cm}
 \subsubsection{Time complexity.} 
 First, suppose that the field arithmetic over $\F$ can be implemented
 using polynomial bits. This covers all the finite fields where the
 field is given by an explicit irreducible polynomial. Also, over any
 such field the polynomial $T(f)$ can be explicitly computed from the
 input arithmetic circuit deterministically in time
 $\poly(L,s,n,d^{O(r)})$.
 
Notice that in each recursive application the number of generators in
the ideal is reduced by at least one. Furthermore, in each recursive
step we need time $\poly(L,s,n,d^{O(r)})$ to run the division
algorithm. This gives us a recurrence of $t(n) \leq t(n-1) +
\poly(L,s,n,d^{{O}(r)})$ which solves to $t(n) \leq
\poly(L,s,n,d^{{O}(r)})$. Over $\Q$, we only need to argue that the
intermediate bit-size complexity growth is only polynomial in the
input size. The proof is given in the appendix (Section \ref{app1})
which involves fairly standard argument. The rest of the argument is
exactly same.
\end{toneproof}
  
\subsection{Vertex Cover Detection in Low Rank Graphs}
In the Vertex Cover problem, we are given a graph $G=(V,E)$ on $n$
vertices and an integer $k$ and the question is to decide whether there is a Vertex Cover of size $k$ in $G$.
This is a classical $\NP$-complete problem. In this section we show an efficient algorithm to detect vertex cover in a graph whose adjacency matrix is of low rank. 

\begin{ttwoproof}
 We present a reduction from Vertex Cover problem to Univariate Ideal Membership problem that produces a polynomial whose 
 rank is almost same as the rank of $A_G$. 
  Consider the ideal $I=\angle{x^2_1 - x_1,x^2_2 - x_2,\ldots, x^2_n - x_n}$ and the polynomial 
 \[f = \prod^{\binom{n}{2}}_{s=1} (\vec{x} A_G  \vec{x}^T - s) \cdot \prod^{n-k-1}_{t=0}\left(\sum^{n}_{i=1} x_i - t\right),\]
 where $A_G$ is the adjacency matrix of the graph $G$ and $\vec{x}=(x_1,x_2,\ldots,x_n)$ is row-vector. 
 
  \begin{lemma}\label{rank-f}
 The rank of the polynomial $f$ is at most $r+1$. 
 \end{lemma}
 
 \begin{proof}
 We note that $A_G$ is symmetric since it encodes an undirected graph. Let $Q$ be an invertible $n \times n$ matrix that diagonalizes $A_G$.
 So we have $Q A_G Q^T = D$ where $D$ is a diagonal matrix with only the first $r$ diagonal elements being non-zero.
 Let $\vec{y}=(y_1,y_2,\ldots,y_n)$ be another row-vector of variables. Now, we show the effect of the transform $\vec{x}\mapsto \vec{y}Q$ on the polynomial
  $\vec{x}A_G \vec{x}^T$.
Clearly, $\vec{y}Q A_G Q^T \vec{y}^T = \vec{y}D\vec{y}^T$ and since there are only $r$ non-zero entries on the diagonal, the polynomial $\vec{y}D\vec{y}^T$ is over the
 variables $y_1,y_2,\ldots,y_r$. Thus $g = \prod^{\binom{n}{2}}_{s=1} (\vec{x}A_G  \vec{x}^T - s)$ is a rank $r$ polynomial. Also $h=\prod^{n-k-1}_{t=0}(\sum^{n}_{i=1} x_i - t)$
 is a rank $1$ polynomial as there is only one linear form $\sum^{n}_{i=1} x_i$. Since $f=gh$, we conclude that $f$ is a rank $r+1$ polynomial.
 \end{proof}
 
 Now the proof of Theorem \ref{lowVC} follows from the next claim. 
 \begin{claim}\label{VC}
  The graph $G$ has a Vertex Cover of size $k$ if and only if $f \not \in I$.
 \end{claim}
 \begin{claimproof}
  First, observe that the set of common zeroes of the generators of the ideal $I$ is the set 
  $\{ 0 ,1 \}^n$. Let $S$ be a vertex cover in $G$ such that  $|S|\leq k$. We will exhibit a point $\vec{\alpha}\in\{0,1\}^n$ such that $f(\vec{\alpha})\neq 0$. This will imply that 
  $f\not\in I$. Identify the vertices of $G$ with $\{1,2,\ldots,n\}$. Define $\vec{\alpha}(i)=0$ if and only if $i\in S$. Since $\vec{x} A_G \vec{x}^T = \sum_{(i,j)\in E_G} x_i x_j$ and $S$ is a vertex cover for $G$, it is clear that $\vec{x} A_G \vec{x}^T(\vec{\alpha})=0$. Also $(\sum_{i=1}^n x_i)(\vec{\alpha})\geq n-k$. Then clearly $f(\vec{\alpha})\neq 0$. 
  
 For the other direction, suppose that $f \not \in I$. Then by Theorem \ref{alon}, there exists 
 $\vec{\alpha}\in\{0,1\}^n$ such that $f(\vec{\alpha})\neq 0$. Define the set $S\subseteq [n]$ as follows. Include $i\in S$ if and only if $\vec{\alpha}(i)=0$.  
 Since $f(\vec{\alpha})\neq 0$, and the range of values that $\vec{x} A_G \vec{x}^T$ can take is $\{0,1,\ldots, |E|\}$, it must be the case that $\vec{x} A_G \vec{x}^T(\vec{\alpha})=0$. It implies that the set $S$ is a vertex cover for $G$.  Moreover, $\prod^{n-k-1}_{t=0}(\sum^{n}_{i=1} x_i - t)(\vec{\alpha})\neq 0$ implies that $|S|\leq k$. 
  \end{claimproof}
  
 The degree of the polynomial $f$ is bounded by $n^2 + n$ and from Claim~\ref{VC} we know that $f \pmod I$ is a non-zero polynomial. By Schwarz-Zippel-Demillo-Lipton~\cite{DL78, Zip79, Sch80} lemma $(f \pmod I)(\vec{\beta})$ is non-zero with high probability when $\vec{\beta}$ is chosen randomly from a small domain. Now using Theorem~\ref{main-thm-4}, we need to just compute $(f \pmod I)(\vec{\beta})$ which can be performed in $(n,k)^{O(r)}$ time.  
 \end{ttwoproof}
\vspace{0cm}

\section{Parameterized Complexity of Univariate Ideals} \label{parameterized-section}\vspace{-.25 cm}

We have already mentioned in Fact~\ref{perm}, that checking if the
integer permanent is zero is reducible to testing membership of a
polynomial $f(\X)$ in the ideal $\angle{x_1^2, \ldots, x_n^2}$.  So
univariate ideal membership is hard for the complexity class
$\cequalp$ even when the ideal is generated by powers of variables
\cite{Sal92}.  In this section we study the univariate ideal
membership with the lens of parametrized complexity. The parameters we
consider are either polynomial degree or number of the generators for
the ideal.

 \vspace{-.25 cm}
\subsection{Parameterized by the Degree of the Polynomial}\vspace{-.15 cm} 
We consider the following: Let $I$ be a univarite ideal given by
generators and $f\in\F[\X]$ a degree $k$ polynomial. Is checking
whether $f$ is in $I$ fixed parameter tractable (with $k$ as
the fixed parameter)?

We show that 
it admits an $\FPT$ algorithm for the special case when
$I=\angle{x_1^{e_1},x_2^{e_2},\ldots,x_n^{e_n}}$.

\subsubsection{Proof of Theorem \ref{degparalgo}}

The proof uses the Hadamard product of polynomials and a
connection to noncommutative computation. This builds on our
recent work \cite{ACDM18}. We include Section \ref{app2} in
the appendix to provide the background. Here, we
recall the Hadamard product of polynomials. Let
$[m]f$ denote the coefficient of the monomial $m$ in the polynomial
$f$. For $f,g \in \F[\X]$, their Hadamard product is defined
as $f \circ g = \sum_{m} [m]f \cdot [m]g \cdot m$. We also
need a slight variant that we call the scaled Hadamard product.
For $f,g \in\F[X]$, their scaled Hadamard Product is
$f \circ^{s} g = \sum_{m} m! \cdot [m]f \cdot [m]g \cdot m$, where
$m=x^{e_1}_{i_1}x^{e_2}_{i_2} \ldots x^{e_r}_{i_r}$ and $m! =e_1!\cdot
e_2!\cdots e_r!$ abusing the notation.

If one of $f,g\in\F[X]$ is multilinear then the scaled Hadamard
product $f\circ^s g$ coincides with the Hadamard product $f\circ g$.

\vspace{.25cm}
\begin{tfourproof}
  The proof consists of following three lemmas. Firstly, given
  an input instance a degree-$k$ $f(\X)$ and ideal
  $I=\angle{x^{e_1}_1,x^{e_2}_2,\ldots,x^{e_n}_n}$ of ideal membership,
  we reduce it to computing the (scaled) Hadamard product of
  $f(\X)$ and a polynomial $g(\X)$, where $g(\X)$ is a
  weighted sum of all degree $k$ monomials that are not in $I$.
    
Then we show that we can compute Hadamard product of any two
polynomials in time roughly linear in the product of the size
of the circuits when one of the polynomials is given by a
diagonal circuit as input. Finally the last part of the proof is
a randomized construction of a homogeneous degree $k$ diagonal
circuit of top fain-in roughly $O^{*}(4.08^k)$ that computes a
polynomial weakly equivalent \footnote{Two polynomials $f$ and $g$
  are said to be weakly equivalent if they share the same set of
  monomials.} to the polynomial $g$ with constant probability.

To define the polynomial $g(\X)$, let $S_{m,k}$ be the elementary
symmetric polynomial of degree $k$ over $m$ variables. Set $m =
\sum_{i=1}^n (e_i - 1)$. Let $S_{m,k}$ is defined over the variable
set $\{z_{1,1},\ldots,z_{1,e_1-1},\ldots,z_{n,1},\ldots,z_{n,e_n-1}\}$. We
define $g(\X)$ as the polynomial obtained from $S_{m,k}$ replacing
each $z_{i,j}$ by $x_i$.

\begin{lemma}\label{lem1-4}
Given integers $e_1,e_2,\ldots,e_n$, and a polynomial $f(\X)$ of degree $k$, $f\in \angle{x^{e_1}_1,x^{e_2}_2,\ldots,x^{e_n}_n}$ if and only if $f\circ^{s} g\equiv 0$.
\end{lemma}
 
\begin{proof}
Suppose, $f\not\in \angle{x^{e_1}_1,x^{e_2}_2,\ldots,x^{e_n}_n}$, then $f$ must contain a degree $k$ monomial $m = x^{f_1}_{1} x^{f_2}_{2}\ldots x^{f_n}_{n}$ such that $f_i<e_i$ for each $1\leq i \leq n$. From the construction, it is clear that $g(\X)$ contains $m$. Therefore, the polynomial $f\circ^{s} g$ is not identically zero. The converse is also true for the similar reason. 
\end{proof}\vspace{-.5 cm}

\begin{lemma}\label{lem2-4}
Given a circuit $C$ of size $s$ computing a polynomial $g \in \F[\X]$ and a homogeneous degree $k$ diagonal circuit $\Sigma\wedge^{[k]}\Sigma$ circuit $D$ of size $s'$ computing $f\in\F[\X]$, we can obtain a circuit computing a polynomial $f\circ^s g$ in deterministic $ss'\cdot\poly(n,k)$ time. Furthermore, for a scalar input $\vec{a}\in\F^n$, we can evaluate $(f\circ^s g)(\vec{a})$ using $\poly(n,k)$ space. 	
\end{lemma} 
The proof easily follows from our recent work \cite{ACDM18}. We include a self-contained proof in the appendix (Section \ref{app2}). 

\begin{lemma}\label{lem3-4}
  There is an efficient randomized algorithm that constructs
  with constant probability a homogeneous degree $k$ diagonal
  circuit $D$ of top fan-in $O^{*}(4.08^k)$ which computes a
  polynomial weakly equivalent to $g$ (defined before
  Lemma~\ref{lem1-4}).
\end{lemma}
 
\begin{proof}
  To construct such a diagonal circuit $D$, we use the idea of
  \cite{Pra18}. We pick a collection of colourings
  $\{\zeta:[m]\to[1.5\cdot k]\}$ of size roughly $O^{*}( (\frac{e}{\sqrt{3}})^k)$ uniformly at random. For each such colouring
  $\zeta_i$, we define a $\Pi^{[1.5\cdot k]}\Sigma$ formula
  $P_i = \prod_{j=1}^{1.5k} (L_j+ 1)$, where $L_j = \sum_{\ell : \zeta_i(\ell) = j} x_\ell$. We say that a monomial is \emph{covered}
  by a coloring $\zeta_i$ if the monomial is in $P_i$. It is easy
  to see that, given any multilinear monomial of degree $k$, the probability that a random coloring will cover the monomial is roughly $(\frac{\sqrt{3}}{e})^k$. 
Hence, going over such a collection of colorings of size $O^{*}((\frac{e}{\sqrt{3}})^k)$ chosen uniformly at random, with a constant probability all the multilinear terms of degree $k$ will be covered.  To take the Hadamard product with a polynomial of degree $k$, we need to extract out the degree $k$ homogeneous part (say $P'_i$) from each $P_i$. Notice that, using elementary symmetric polynomial over $1.5k$ many variables $S_{1.5k,k}$, we can write $P'_i = S_{1.5k,k}(L_1,\ldots,L_{1.5k})$.
Now we use Lemma \ref{Lee} to get a diagonal $\Sigma\wedge^{[k]}\Sigma$ circuit of top fan-in roughly $\binom{1.5k}{0.5k}$ for each $P'_i$. Define $D = \sum_{i=1}^{O^{*}((\frac{e}{\sqrt{3}})^k)}P'_i$. By a direct calculation, one can obtain a diagonal circuit $D$ of top fan-in $O^{*}(4.08^k)$ 
which is weakly equivalent to the polynomial $S_{m,k}$. 
The construction of the polynomial $g(\X)$ from $S_{m,k}$ is already explained before Lemma \ref{lem1-4}. 
\end{proof}

\vspace{-.2cm}
Now, given a circuit $C$ computing $f\in \F[\X]$ and integers $e_1,\ldots,e_n$, to decide the membership of $f$ in the ideal $I=\angle{x^{e_1}_1,\ldots,x^{e_n}_n}$, we construct a diagonal circuit $D$ from Lemma~\ref{lem3-4} and take (scaled) Hadamard product with $C$ using Lemma~\ref{lem2-4}. Following Lemma~\ref{lem1-4}, we can decide the membership of $f$ in the ideal checking the polynomial $C\circ^s D$ is identically zero or not which can be performed by random substitution using Schwartz-Zippel Lemma \cite{Sch80, Zip79}. Over $\Z$ the given circuit can compute numbers as large as $2^{2^{n^{O(1)}}}$. To handle this while we evaluate the circuit, we do the evaluation modulo a random polynomial bit prime. This is a standard idea.  
\end{tfourproof}

\vspace{-.2cm}
\subsection{Parameterized by Number of Generators}\label{hardness}

In this section, we consider the univariate ideal membership
parameterized on the number of generators of the ideal. More
precisely, given a polynomial $f(\X)$, can we obtain an $\FPT$
algorithm for testing membership in the univariate ideal
$\angle{p_1(x_1), \ldots, p_k(x_k)}$ parameterized by $k$?  We show
that the problem is $\W[1]$-hard. Moreover, in contrast to the
previous case, we obtain $\MINI$-hardness for a special case of the
problem when the univariate generators are just power of variables.

\begin{tfiveproof}
We show a reduction from \emph{$k$-independent set}, a well known $\W[1]$-hard problem \cite{FKLMPPS15}, to this problem. Let $G=(V,E)$ be a graph on $n$ vertices and $k$ be the size of the independent set.  We identify its vertex set with the numbers $\{1,2,\ldots,n \}$ and the edges are tuples over 
$[n] \times [n]$. Define the univariate ideal $I=\angle{p_1(x_1), \ldots, p_k(x_k)}$ where for each $1\leq i \leq k$, we define $p_i(x_i) = \prod^{n}_{j=1}(x_i - j)$.  Now we are going to define a polynomial $f$ that uses only $k$ variables which will be used for the ideal membership problem.  
First consider the polynomial $D= \prod_{1\leq i \neq j\leq k}(x_i - x_j)$. 
 
Now we define the polynomial, 
\[f= \prod_{1\leq i \neq j\leq k} \prod_{(u,v) \in E\subseteq [n]\times [n]} [(x_i - u)^2 + (x_j - v)^2] \cdot [(x_j - u)^2 + (x_i - v)^2 ].\]
The proof follows from the following claim.  
\begin{claim}
  $f\cdot D \not \in \angle{p_1(x_1),p_2(x_2),\dots,p_k(x_k)}$ if and only if $G$ has an independent set of size $k$.
 \end{claim}
 \begin{claimproof}
  We use Theorem \ref{alon} to prove the claim. Let $\{ j_1,j_2,\ldots,j_k \}$ be an independent set in $G$. Notice that $(j_1, \ldots, j_k)$ is a common zero of the generators $p_1, \ldots, p_k$. 
  Now notice that $f \cdot D$ does not vanish at the point $(j_1, \ldots, j_k)$ as all the edges 
  $(j_{\ell},j_{\ell'}) : 1\leq \ell, \ell'\leq k$ are absent in the edge set $E$. 
  Thus there is a common root of the ideal on which $f \cdot D$ does not vanish and hence $f\cdot D \not \in \angle{p_1(x_1),p_2(x_2),\dots,p_k(x_k)}$.
  
  Now if $f\cdot D \not \in \angle{p_1(x_1),p_2(x_2),\dots,p_k(x_k)}$ then there is a common zero 
  $(j_1, \ldots,j_k)$ of the ideal on which $f \cdot D$ does not vanish. Using the same argument one can easily see that $\{j_1, \ldots, j_k\}$ is an independent set in $G$. 
   \end{claimproof}
\end{tfiveproof}

\vspace{0cm}
\subsubsection{Proof of Theorem \ref{genparuna}}

We first show a reduction from the linear algebraic problem $\kLINEQ$
to our univariate ideal membership problem.
\begin{definition}\textbf{$\kLINEQ$}\\
\textit{Input:} Integers $k,n$ in unary, a $k\times n$ matrix $A$ with
all the entries given in unary and a $k$ dimensional vector $\vec{b}$
with all entries in unary.\\ \textit{Parameter:
  k}.\\ \textit{Question:} Does there exist an $\vec{x}\in \{0,1\}^n$
such that $A\vec{x} = \vec{b}$?
\end{definition}

 It turns out that $\kLINEQ$ problem is more amenable to the
 $\MINI$-hardness proof. Finally we show a reduction from $\motsat$ to
 $\kLINEQ$ to complete the proof. It  is easy to observe from the standard \textit{Schaefer Reduction} \cite{Sch78} that $\motsat$ is $\MINI$-hard. The full proof is given in the appendix (Section \ref{app2}).

 \section{Non-deterministic Algorithm for Univariate Ideal Membership}\label{main-thm1-sec}
 In this section we prove Theorem \ref{main-thm-1}. 
 Given a polynomial $f(\X)\in\Q[\X]$ and a univariate ideal $I=\angle{p_1(x_1), \ldots, p_n(x_n)}$ where the generators are $p_1, \ldots, p_n$, we show a non-deterministic algorithm to decide the 
 (non)-membership of $f$ in $I$. By Theorem \ref{alon}, it suffices to show that there is a common zero $\vec{\alpha}$ of the generators $p_1, p_2, \ldots, p_n$ such that $f(\alpha)\neq 0$. 
 Since in general $\vec{\alpha}\in\C^n$, it is not immediately clear how to guess such a common zero by a $\NP$ machine. However, we are able to show that for the $\NP$ machine it suffices to guess such an $\vec{\alpha}$ upto polynomially many bits of approximation. 
 
 We begin by proving a few technical facts which are useful for the main proof. 
 Write $f(\X)= \sum_{i=1}^n h_i(\X) ~p_i(x_i) + R(\X)$ where for all $i\in [n]$, $\deg_{x_i} (R) < \deg(p_i)$. For any polynomial $g$,  let $|c(g)|$ be the maximum coefficient (in absolute value) appearing in $g$. The following lemma gives an estimate for the coefficients of the polynomials $h_1, \ldots, h_n, R$.   
\vspace{-.1cm}
\begin{lemma}\label{lemma-3}
Let $2^{-L}\leq |c(f)|, |c(p_i)|\leq 2^L$. Then there is $L'=\poly(L, d, n)$ such that $2^{-L'}\leq |c(h_i)|, |c(R)|\leq 2^{L'}$ where $d$ is the degree upper bound for $f$, and $\{p_i : 1\leq i\leq n\}$. 
\end{lemma} 

\begin{proof}
The estimate on $L'$ follows implicitly from the known results~\cite{coll67}. 
It can be also seen by direct computation.   
Write $f(\X) = \sum_{i} f_i(x_2, \ldots, x_n) ~x_1^i$ and then divide $x_1 ^i \pmod{p_1(x_1)}$ for each $i$. The modulo computation can be done by writing $x_1^i = q_1(x_1) p(x_1) + r_1(x_1)$ with the coefficients of $q_1$ and $r_1$ are unknown. We can then solve it using standard linear algebra. In particular, one can use the Cramer's rule for system of linear equation solution. The growth of the bit-size is only $\poly(L,d)$. More precisely, if $c_{\max}$ is the maximum among 
$|c(f)|, |c(p_1)|$, any final coefficient is at most $c_{\max}\cdot 2^{\poly(L,d)}$.  We repeat the  procedure for the other univariate polynomials one by one. The final growth on the coefficients size is 
at most $\poly(n,L,d)$. 
\end{proof}

\vspace{-.05cm}

Let $\vec  \alpha = (\alpha_1, \ldots, \alpha_n) \in \C^n$ be such that $p_i(\alpha_i) = 0$, $1\leq i\leq n$. 
From Lemma~\ref{lemma-1}, we get that $\frac{1}{2^{\hat{L}}} \leq |\alpha_i|\leq 2^{\hat{L}}$ where 
$\hat{L} = \poly(L,d)$. Let $\tilde{\alpha}_i\in\Q[i]$ be an $\epsilon$-approximation of $\alpha_i$, e.g. $|\alpha_i-\tilde{\alpha}_i|\leq \epsilon$. Then we show that the absolute value of $p_i(\tilde{\alpha}_i)$ is not too far from zero. 
\vspace{-.05cm}
\begin{observation}\label{obs-1}
For $1\leq i\leq n$ we have that $|p_i(\tilde{\alpha}_i)|\leq \epsilon \cdot 2^{(d L)^{O(1)}}$. 
\end{observation}

\begin{proof}
Let $p_i(x_i) = c\cdot \prod_{j=1}^d  (x_i - \beta_{i,j})$ and w.l.o.g assume that $\tilde{\alpha}_i$ is the approximation of the root $\beta_{i,1}$. Then $|p_i(\tilde{\alpha}_i)| \leq \epsilon \cdot |c|\cdot \prod_{j=2}^d 
|\tilde{\alpha}_i - \beta_{i,j}| \leq \epsilon \cdot |c|\cdot \prod_{j=2}^d 
(|\beta_{i,1} - \beta_{i,j}| + \epsilon)\leq \epsilon \cdot 2^{\poly(d, L)}$.  The final bound follows from the bound on the roots given in Lemma \ref{lemma-1}. 
\end{proof}

Since we have an upper bound on the coefficients of the polynomials $\{h_i : 1\leq i\leq n\}$  from Lemma ~\ref{lemma-3}, it follows that for $1\leq i\leq n$ we have that $|h_i(\tilde{\alpha})|\leq 2^{(n d L)^{O(1)}}$. Here we use the fact that the approximate root ${\alpha}_i$ can be trivially bounded by $2^{\hat{L}+1}$.   

\vspace{-0.05cm}

 \begin{tsevenproof}
 If $f$ is not in the ideal $I$, by Alon's Nullstellensatz, we know that there exists a tuple $\vec\alpha = (\alpha_1, \ldots, \alpha_n) \in Z(p_1) \times \ldots \times Z(p_n)$ such that 
$R(\vec\alpha) \neq 0$. 
Suppose that the $\NP$ Machine guess the tuple $\vec{\tilde{\alpha}}=(\tilde{\alpha}_1, \ldots, \tilde{\alpha}_n)$ which is the $\epsilon$-approximation of the tuple $\vec \alpha=(\alpha_1, \ldots, \alpha_n)$. Using the black-box for $f$, obtain the value for $f(\vec{\tilde{\alpha}})$. Next, we show that 
the value $|f(\vec{\tilde{\alpha}})|$ distinguishes between the cases $f\in I$ and $f\not\in I$.  The full proof is given in the appendix (Section \ref{app3}). The proof uses Lemma \ref{lemma-3} and Observation \ref{obs-1}.
If $f\in I$, we show that  $|f(\vec{\tilde{\alpha}})| \leq \epsilon\cdot 2^{(n d L)^{c_2}}$. where the constant $c_2$ is fixed by Observation \ref{obs-1} and the bounds on $|h_i(\vec{\tilde{\alpha}})|$.  
If $f\not\in I$, we show that $|f(\vec{\tilde{\alpha}})| \geq \frac{1}{2^{(n d L)^{c_3}}} - \epsilon \cdot (2^{(n d L)^{c_4}} + 2^{(n d L)^{c_2}})$, for some constant $c_3$ and $c_4$.  
To make the calculation precise, let $3M = \frac{1}{2^{(n d L)^{c_3}}}$ and choose 
$\epsilon$ such that $\epsilon \cdot (2^{(n d L)^{c_4}} + 2^{(n d L)^{c_2}}) \leq M$. 
The final implication will be $|f(\vec{\tilde{\alpha}})|\leq M$ when $f\in I$ and $|f(\vec{\tilde{\alpha}})|\geq 2M$ when $f\not\in I$. It is important to note that the parameter $M$ can be pre-computed from the input parameters efficiently. 

\end{tsevenproof}

\bibliographystyle{alpha}

\appendix

\section{Proof of Theorem \ref{main-thm-4}}\label{app1}
\subsection{Lemmas for the proof of Theorem \ref{main-thm-4}}

\begin{tloneproof}
From the uniqueness of the remainder for the univariate ideals, we get that $R(\X) = R_r(\X) \pmod{I_{[n]\setminus[r]}}$. Now we write explicitly the polynomial $R_r(\X)$ as 
$R_r= \sum_{\bar{u}} r_{\bar{u}} \cdot x_{r+1}^{u_1}\ldots, x_n^{u_{n-r}}$ where 
$r_u\in\F[\X_{[r]}]$. 
So we get that, 
\[
R_r \pmod{I_{[n]\setminus[r]} }= \sum_{\bar{u}} r_{\bar{u}} \prod_{j=1}^{n-r} q(x_{r+j})
\]
where $q(x_{r+j}) = x^{u_j}_{r+j}\pmod{p(x_{r+j})}$. 
Then the lemma follows by substituting $x_1=\alpha_1, \ldots, x_r = \alpha_r$ in the 
relation $R = R_r \pmod{I_{[n]\setminus[r]}}$.  
\end{tloneproof}




 
 \begin{tltwoproof}
 Let $f = \sum_{i=1}^r h_i(\X) \cdot p_i(x_i) + R(\X)$ and $T(f) = \sum_{i=1}^r h'_i(\X) \cdot p_i(x_i) + R'(\X)$. Note that 
 $\deg_{x_i}R,  \deg_{x_i}R'< \deg(p_i(x_i))$ for $1\leq i\leq r$. Since $T$ is invertible and also fixes 
 $x_1, \ldots, x_r$, we can write $f = \sum_{i=1}^r T^{-1}(h'_i(\X)) \cdot p_i(x_i) + T^{-1}(R'(\X))$. By the property of $T$ it is clear that $\deg_{x_i}(T^{-1}(R'(\X))) < \deg(p_i(x_i))$ for $1\leq i\leq r$. Combining two expression for $f$, we immediately conclude that $(R - T^{-1}(R')) = 0 \pmod{I_{[r]}}$ which forces that $R = T^{-1}(R')$.   
 \end{tltwoproof}

\subsection{Bit-size growth over $\mathbb{Q}$ for Theorem \ref{main-thm-4}}
Let $\tilde{L}$ be the maximum bit size of any coefficient appearing in $f(z_1,\ldots,z_r)$, and
let $L$ be an upper bound on the bit sizes of the other inputs, i.e.\ bit sizes of coefficients of $\ell_1,\ldots,\ell_r, p_1,\ldots,p_n$ and $\alpha_1,\ldots,\alpha_n$.
We will show that the circuit that we use in the next recursive step has coefficients of bit size at most $\tilde{L} + \poly(n,d,L)$.

Let $|c(h)|$ denote the maximum coefficient (in absolute value) appearing in any polynomial $h$. Then by direct expansion we can see that $|c(f(\ell_1,\ldots, \ell_r))| \leq 2^{\tilde{L} + \poly(n,d,L)}$. Also the linear transformation from lemma~\ref{varsep} can be implemented using $\poly$-bit size entries. 
Together,  we get that that $c(T(f(\ell_1,\ldots,\ell_r)) \leq 2^{\tilde{L} + \poly(n,d,L)}$. At this point, we expand the circuit and obtain $T(f)$ explicitly as a sum of $d^{O(r)}$ monomials. Then divide $T(f)$ by $p_1(x_1),\ldots,p_r(x_r)$ one-by-one, and
substitute $x_1=\alpha_1,\ldots,x_r=\alpha_r$ giving us the remainder $g(x_{r+1},\ldots,x_{r+r'})$. We note that $|c(g)| \leq 2^{\tilde{L} + \poly(n,d,L)}$ \footnote{We tackle a similar situation in Section \ref{main-thm1-sec}, and Lemma \ref{lemma-3} gives further explanation on the bit-complexity growth when we divide by univariate polynomials.}.
Now the algorithm passes the $d^{O(r)}$ size $\Sigma \Pi \Sigma$ circuit $g(\ell_{1,2},\ldots,\ell_{r' , 2})$ (We note that $T^{-1}(x_{r+1})=\ell_{1,2},\ldots,T^{-1}(x_{r+r'})=\ell_{r' , 2}$), univariates $p_{r+1}(x_{r+1}),\ldots,p_n(x_n)$ and the point $(\alpha_{r+1},\ldots,\alpha_n)$
for the next recursive call. 

We note that the bit-size upper bound $L$ does not change for the input linear forms, and the coefficient bit-size of $f$ grows from
$\tilde{L}$ to $\tilde{L} + \poly(n,d,L)$ in one step of the recursion. This gives us the recurrence $S(n) \leq S(n-1) + \poly(n,d,L)$ with 
$S(1)=\tilde{L}$. Which solves to $S(n) = O(\tilde{L} + \poly(n,d,L))$. 

\section{Proofs in Section \ref{parameterized-section}}\label{app2}
\subsection{Background for proof of Theorem \ref{degparalgo}}\label{background-degparalgo}

\subsection*{Hadamard Product}
We recall the definition of Hadamard product of two polynomials. 

\begin{definition} \label{hadprod}
 Given two polynomials $f,g \in \F[\X]$,  the Hadamard product $f \circ g$ is defined as
  $f \circ g = \sum_{m} [m]f \cdot [m]g \cdot m$.
 \end{definition}

In this paper we adapt the notion of Hadamard product suitably and define a scaled version of Hadamard Product of two polynomials. 

\begin{definition} \label{s-hadprod} Given two polynomials $f,g \in
  \F[X]$, their \emph{scaled Hadamard Product}
 $f\circ^{s} g$, is defined as
$f \circ^{s} g = \sum_{m} m! \cdot [m]f \cdot [m]g \cdot m$,
where $m=x^{e_1}_{i_1}x^{e_2}_{i_2} \ldots x^{e_r}_{i_r}$ and 
$m! =e_1!\cdot e_2!\cdots e_r!$ abusing the notation.
\end{definition}

\begin{remark}
  Given two polynomials $f\in\F[X]$ and $g\in\F[X]$, if one of these
  two is a multilinear polynomial then scaled Hadamard product
  $f\circ^s g$ is same as Hadamard product $f\circ g$.
\end{remark}

\subsubsection*{Connection to noncommutative computation}

In this paper, we will also deal with the free noncommutative ring
$\F\angle{Y}$, where $Y$ is a set of noncommuting variables. 
Given a commutative circuit $C$
computing a polynomial in $\F[x_1,x_2,\ldots,x_n]$, the  \emph{noncommutative
version} of $C$, $C^{nc}$
as the noncommutative circuit obtained from $C$ by
fixing an ordering of the inputs to each product gate in $C$ and replacing $x_i$
by the noncommuting variable $y_i, 1\le i\le n$.
Thus, $C^{nc}$
will compute a polynomial $f^{nc}_{C}$ in the ring $\F\angle{Y}$,
where $Y=\{y_1,y_2,\ldots,y_n\}$ are $n$ noncommuting variables.

\subsubsection*{Symmetric polynomial and weakly equivalent polynomial}
The symmetric polynomial of degree $k$ over $n$ variables $\{x_1,x_2,\ldots,x_n\}$, denoted by $S_{n,k}$, is defined as follows:
$S_{n,k}(x_1,x_2,\ldots , x_n) = \sum_{T\subseteq [n],|T| = k} \prod_{i\in T}x_i$.
Notice that, $S_{n,k}$ contains all the degree $k$ multillinear terms. 
A recent result of Lee gives the following homogeneous diagonal circuit for $S_{n,k}$ \cite{Lee15}. 

\begin{lemma}\label{Lee}
The symmetric polynomial $S_{n,k}$ can be computed by a homogenous $\Sigma^{[s]}\wedge^{[k]}\Sigma$ circuit where $s\leq \sum_{i=0}^{k/2} \binom{n}{i}$. 
\end{lemma}
A polynomial $f\in\F[X]$ is said to be weakly equivalent to a polynomial $g\in\F[X]$, if the following is true. For each monomial $m$, $[m]f = 0$ if and only if $[m]g = 0$. Moreover, if $[m]f\geq 0$ for each monomial $f$,  we define $f$ to be a positively weakly equivalent polynomial to $g$.
One can define the same in noncommutative setting also.
 In this paper, we will use polynomials weakly equivalent to $S_{n,k}$.

\vspace{-.25cm}

\subsection{The proof of Lemma \ref{lem2-4}}

As (scaled) Hadamard product distributes over addition, it is sufficient to prove the lemma for each $\wedge^{[k]}\Sigma$ sub-circuits. Fix a   $\wedge^{[k]}\Sigma$ sub-circuit $D'$. Our goal is to compute $C\circ^s D'$ efficiently. By the distributivity property it follows that the final running time will be at most $s'$ times the time taken for computing the scaled Hadamard product with any such sub-circuit.  
Let us consider the noncommutative version of $D'$, $D'^{nc}$ computing noncommutaive polynomial $\hat{f}\in \F\angle Y$.
 Let $X_k$ denote the set of all degree $k$ monomials over $X$. Also, $Y^k$ denotes all degree $k$ noncommutative monomials (i.e., words) over $Y$. Each monomial $m \in X_k$ can appear as different noncommutative words $\hat{m}$ in $\hat{f}$.
We use the notation $\hat{m}\to m$ to denote that $\hat{m} \in Y^k$ will be transformed to $m \in X_k$ by substituting $x_i$ for $y_i$, $1 \leq i \leq n$. Then, we observe that 
\[[m]f = \sum_{\hat{m}\to m}[\hat{m}]\hat{f}.\]
Moreover, a $\wedge^{[k]}\Sigma$ circuit has the following useful property. For each pair $\hat{m},m'$ such that $\hat{m}\to m$ and $m'\to m$, $[\hat{m}]\hat{f} = [m']\hat{f}$.
Now, we want to bound the number of words $\hat{m}$ such that $\hat{m}\to m$ for each monomial $m$. It is easy to see that for each monomial $m$, there are $\frac{k!}{m!}$ such noncommutative words. Therefore, \[[\hat{m}]\hat{f} = \frac{m!}{k!}\cdot [m]f.\]

We consider the noncommutative version of $C$, $C^{nc}$ and note that, $D'^{nc}$ has a small ABP. Therefore, using the result of~\cite{AJS09}, we can compute $C^{nc}\circ D'^{nc}$ in $\poly(|C|, |D'|)$ time. Let us denote $\tilde{C}$ as the commutative version of this circuit.
Suppose, $f = \sum [m]f\cdot m$. 
Hence, for each monomial $m\in X_k$,
\begin{align*}
[m]\tilde{C} &= \sum_{\hat{m}\to m} [\hat{m}](C^{nc}\circ D'^{nc}) \\
&= \sum_{\hat{m}\to m} [\hat{m}]C^{nc} \cdot [\hat{m}]D'^{nc}\\ 
&= \sum_{\hat{m}\to m} [\hat{m}]C^{nc} \cdot \frac{m!}{k!}\cdot [m]f\\
&= \frac{m!}{k!}\cdot [m]f \cdot \sum_{\hat{m}\to m} [\hat{m}]C^{nc}\\
&= \frac{m!}{k!}\cdot [m]f \cdot [m]g.       
\end{align*}
Therefore $k!\cdot \tilde{C}$ computes the scaled Hadamard product of $f$ and $g$, that proves the first part of the theorem. To prove the second part, notice that, given a scalar $\vec{a}\in \F^n$, we can compute the commutative scaled Hadamard product of $g$ and each $\wedge^{[k]}\Sigma$ sub-circuit and evaluate it at $\vec{a}$. Hence, $(f\circ^s g)(\vec{a})$ can be computed incrementally using only $\poly(n,k)$ space.

\subsubsection{Proof of Theorem \ref{genparuna}}

We first relate our univariate ideal membership problem with a linear algebraic problem $\kLINEQ$. It turns that $\kLINEQ$ problem is more amenable to the $\MINI$-hardness proof. Finally we show a reduction from $\motsat$ to $\kLINEQ$ to complete the proof.  

\begin{definition}\textbf{$\kLINEQ$}\\
\textit{Input:} Integers $k,n$ in unary, a $k\times n$ matrix $A$ with all the  entries given in unary and a $k$ dimensional vector $\vec{b}$ with all  entries in unary.\\
\textit{Parameter: k}.\\
\textit{Question:} Does there exist an $\vec{x}\in \{0,1\}^n$ such that $A\vec{x} = \vec{b}$? 
\end{definition}

\begin{lemma}\label{k-lin-to-im}
There is a parameterized reduction from $\kLINEQ$ to the univariate ideal membership problem when the ideal is given by the powers of variables as generators.
\end{lemma}
\begin{proof}
We introduce $2k$ variables $x_1,x_2,\dots,x_k,y_1,y_2,\dots,y_k$ where two variables will be used for each row. For each $i\in [n]$, let $\mu_i = \sum_{j=1}^n a_{ij}$. For each column $c_i = (a_{1i},a_{2i},\dots,a_{ki})$ we construct the polynomial
$P_i = ({y_1}^{a_{1i}}{y_2}^{a_{2i}}\dots {y_k}^{a_{ki}} + {x_1}^{a_{1i}}{x_2}^{a_{2i}}\dots {x_k}^{a_{ki}})$. We let $P_A = \prod_{i=1}^{n} P_i$ and we choose the ideal to be 
$\langle x_1^{b_1 + 1},y_1^{\mu_1 - b_1 +1},$ $\dots,x_k^{b_k + 1},y_1^{\mu_k - b_k +1}\rangle $. Notice that $P_A$ has a small arithmetic circuit which is polynomial time computable.

\begin{claim}\label{bin-general}
An instance $
(A,\vec{b})$ is an YES instance for $\kLINEQ$ iff $P_A \not \in \langle x_1^{b_1 + 1},y_1^{\mu_1 - b_1 +1},\dots,x_k^{b_k + 1},y_k^{\mu_k - b_k +1}\rangle $.
 \end{claim}

\begin{proof}
Suppose $(A,\vec{b})$ is an YES instance. Then there is an $\vec{x}\in \{0,1\}^n$ such that $A\vec{x}=\vec{b}$. Define $S:=\{i\in[n] :  \vec{x}_i=1\}$ where $\bf{x}_i$ is the $i$th co-ordinate of $\vec{
 x}$. Think of the monomial where ${x_1}^{a_{1i}}{x_2}^{a_{2i}}\dots{x_k}^{a_{ki}}$ is picked from $P_i$ for each $i\in S$ and ${y_1}^{a_{1i}}{y_2}^{a_{2i}}\dots {y_k}^{a_{ki}}$ is picked from reaming $P_j$'s where $j\in \bar{S}$.  This gives us the monomial $x_1^{b_1} y_1^{\mu_1 - b_1} \ldots x_k^{b_k} y_1^{\mu_k - b_k}$ in the polynomial $P_A$. Thus $P_A \not \in \angle{x_1^{b_1 + 1},y_1^{\mu_1 - b_1 +1},\ldots,x_k^{b_k + 1},y_k^{\mu_k - b_k +1}} $.
 
 Now we show the other direction. 
 Now suppose $P_A \not \in \langle x_1^{b_1 + 1},y_1^{\mu_1 - b_1 +1},\dots,x_k^{b_k + 1},y_k^{\mu_k - b_k +1}\rangle $.
Let $S := \{ i\in[n] :  {x_1}^{a_{1i}}{x_2}^{a_{2i}}\dots{x_k}^{a_{ki}}$ is picked from  $P_i \}$. There must be a monomial ${x_1}^{c_1}{x_2}^{c_2}\dots{x_k}^{c_k} {y_1}^{d_1}{y_2}^{d_2}\dots{y_k}^{d_k}$ in $P_A$ such that for each $i$, $\sum_{j\in S}a_{ij}=c_i \leq b_i$ , $\sum_{j\not \in S}a_{ij} = d_i \leq( \mu_i - b_i) $.  As, $\mu_i = \sum_{j\in S} a_{ij} + \sum_{i\not \in S}a_{ij}$, we get $b_i \leq \sum_{j\in S}a_{ij}$. Hence, $\sum_{j \in S}a_{ij} = b_i$ for each $i$. Define $\vec{x}\in\{0,1\}^n$ where $\vec{x}_i = 1$  if $i\in S$ else $\vec{x}_i=0$. This shows  $(A,\vec{b})$ is an YES instance. \end{proof}

 \end{proof}
 
 Before we prove the $\MINI$-hardness of $\kLINEQ$, we show that the following problem is 
 $\MINI$-hard.
\begin{definition}
\textbf{$\motsat$}\\
\textit{Input:} Integers $k,n$ in unary, a 3-SAT instance $\mathcal{E}$ consisting of only positive literals where $\mathcal{E}$ has at most $k\log n $ variables and atmost $k\log n$ clauses.\\
\textit{Parameter: k}.\\
\textit{Question:} Does there exist a satisfiable assignment for $\mathcal{E}$ such that every clause has exactly one $\TRUE$ literal? 
\end{definition}

\begin{claim} \label{cl1}
$\motsat$  is $\MINI$-hard.
\end{claim}
To prove the claim we only need to observe that 
the standard \textit{Schaefer Reduction} \cite{Sch78} from 3-SAT to  $\otsat$ is in fact a linear size reduction, that directly gives us an $\FPT$ reduction from $\msat$ to $\motsat$.

\begin{tsixproof}
Given a $\motsat$ instance $\mathcal{E}$, order the variables $v_1,\dots,v_{k\log n}$ and the clauses $C_1,\dots,C_{k\log n}$. Construct the following $k\log n\times k\log n$ matrix $M$ where the rows are indexed by the clauses and the columns are indexed by the variables.  $M[i][j]$ is set to 1 if $v_j$ appears in $C_i$, otherwise set it to 0.  Make $M$ a $2k\log n\times n$ matrix by adding an all zero row between every rows and appending all zero columns at the end. Now, define $\vec{e}$ as a $2k\log n$ dimensional vector where $i$th co-ordinate of $e$, $e_i = 1$ when $i$ is odd and $e_i = 0$ when $i$ is even.
We want to find $\vec{y}\in \{0,1\}^n$ such that $M\vec{y}=\vec{e}$.

However this is not an instance of $\kLINEQ$. To make it so, we observe that $M$ is a bit matrix and $\vec{e}$ is a bit vector, hence we can modify them to a $k\times n$ matrix $A$ and $k$ dimensional vector $\vec{b}$ in the following way. For each column $j$, think of the $i$th consecutive $2\log n$ bits  as the binary expansion of a single entry, call it $N$ and set  $A[i][j]$ to $N$.  Similarly, we modify $\vec{e}$ to a $k$ dimensional vector $\vec{b}$ by considering $2\log n$ bits as a binary expansion of a single entry. Now the proof follows from the following claim.

\begin{claim} \label{cl2}
 $\mathcal{E}$ is an YES instance for $\motsat$ if and only if there exists an $\vec{x}\in \{0,1\}^n$ such that $A\vec{x} = \vec{b}$.
\end{claim}

\begin{proof}
Suppose there is such a satisfiable assignment for $\mathcal{E}$.  Define $S:=\{j\in[k\log n]\mid v_j=\TRUE\}$. Define $\vec{z}\in\{0,1\}^n$ such that $z_j = 1$ where $j\in S$ else $z_j = 0.$ For each $i$, as $C_i$ contains exactly one $\TRUE$ literal, hence $e_{2i+1} = \sum_{j=1}^n M[i][j]\cdot z_j = 1$ and $e_{2i} = 0$. Therefore $\vec{z}$ is a solution for $M\vec{y} = \vec{e}$. 
As every integer has a unique binary expansion, hence $\vec{z}$ is also a solution for $A\vec{x} = \vec{b}$.

Now we prove the other direction. Suppose $A\vec{z} = \vec{b}$ for some $\vec{z}\in\{0,1\}^n$. From the construction of the matrix $M$, it is sufficient to show that $\vec{z}$ is a satisfying assignment for $M\vec{y} = \vec{e}$. 
First we note that the numbers $A[i][j],b[i]$ in their binary expansion have bits 1 in the odd location and 0 in the even locations.
 Let $A[i][j] = \sum^{2\log n}_{t=1} a_{ijt} 2^{t-1}$ and $b[i]=\sum^{2\log n}_{t=1} e_t 2^{t-1}$. 
 Since $A\vec{z} = \vec{b}$ we have $\sum_{j=1}^n A[i][j]\cdot z_j = b[i]$.
 This shows that 
 \begin{equation}
  \begin{split}
   \sum_{j=1}^n A[i][j]\cdot z_j &= \sum_{j=1}^n \left(\sum^{2\log n}_{t=1} a_{ijt} 2^{t-1}\right) \cdot z_j \\
   &=\sum^{2\log n}_{t=1} \left(\sum_{j=1}^n  a_{ijt} \cdot z_j\right) 2^{t-1}.
  \end{split}
 \end{equation}

Since $\mathcal{E}$ is a 3-CNF formula we have $(\sum_{j=1}^n  a_{ijt} \cdot z_j) \in \{0,1,2,3\}$.
Now we compare $(\sum_{j=1}^n  a_{ijt} \cdot z_j)$ with the binary expansion of $b[i]$. 
When $t$ is odd the bit $e_t$ is 1 and so there must be a 1 in the corresponding bit
 of $(\sum_{j=1}^n  a_{ijt} \cdot z_j)$. This shows that $(\sum_{j=1}^n  a_{ijt} \cdot z_j) \neq 0$ when $t$ is odd.
 Now if $(\sum_{j=1}^n  a_{ijt} \cdot z_j) \in \{ 2,3 \} $ for any odd $t$ then the term $2^{t+1}$ will
be produced and this will not match the expansion of $b[i]$ as the $e_{t+1}=0$. Thus by the uniqueness of
binary expansion we conclude that $(\sum_{j=1}^n  a_{ijt} \cdot z_j) = 1$ if $t$ is odd and $0$ otherwise.
Thus $M\vec{y} = \vec{e}$ has a solution with $y_i = z_i$.
\end{proof}

\end{tsixproof}

\section{Proof of Theorem \ref{main-thm-4}}\label{app3}
\subsection{Proof of Theorem \ref{main-thm-1}}  
 \begin{proof}
 If $f$ is not in the ideal $I$, by Alon's Nullstellensatz, we know that there exists a tuple $\vec\alpha = (\alpha_1, \ldots, \alpha_n) \in Z(p_1) \times \ldots \times Z(p_n)$ such that 
$R(\vec\alpha) \neq 0$. 
Suppose that the $\NP$ Machine guess the tuple $\vec{\tilde{\alpha}}=(\tilde{\alpha}_1, \ldots, \tilde{\alpha}_n)$ which is the $\epsilon$-approximation of the tuple $\vec \alpha=(\alpha_1, \ldots, \alpha_n)$  \footnote{Later we fix $\epsilon$ suitably and use Lemma \ref{npguess} to verify in polynomial time that $\vec{\tilde{\alpha}}$ is indeed $\epsilon$-approximation of $\vec\alpha$.}.  Using the black-box for $f$, obtain the value for $f(\vec{\tilde{\alpha}})$. Next, we show that 
the value $|f(\vec{\tilde{\alpha}})|$ distinguishes between the cases $f\in I$ and $f\not\in I$.  \\

\textbf{Case 1 :  $f\in I$} 

$|f(\vec{\tilde{\alpha}})| = |\sum_{i=1}^n h_i(\vec{\tilde{\alpha}}) p_i(\tilde{\alpha_i})|\leq (\sum_{i=1}^n |h_i(\vec{\tilde{\alpha}})|) \cdot \epsilon \cdot 2^{(d L)^{c_1}} \leq \epsilon\cdot 2^{(n d L)^{c_2}}$. where the constant $c_2$ is fixed by Observation \ref{obs-1} and the bounds on $|h_i(\vec{\tilde{\alpha}})|$.  \\

\textbf{Case 2 : $f\not\in I$} 

Recall the inequality for complex numbers  :  $|Z_1 + Z_2| \geq |Z_2| - |Z_1|$. Using this write 
$|f(\vec{\tilde{\alpha}})| \geq |R(\vec{\tilde{\alpha}})| -  \sum_{i=1}^n |h_i(\vec{\tilde{\alpha}})|~|p_i(\vec{\tilde{\alpha}})|$.  
Notice that $|R(\vec{\tilde{\alpha}})| \geq |R(\vec\alpha)| - |R(\vec{\tilde{\alpha}}) - R(\vec\alpha)|$. Combining we get the following : 
$|f(\vec{\tilde{\alpha}}) \geq |R(\vec\alpha)| - |R(\vec{\tilde{\alpha}}) - R(\vec\alpha)| -\epsilon \cdot 2^{(n d L)^{c_2}}. 
$

Now to complete the proof, we show a lower bound on $|R(\vec\alpha)|$ and an upper bound for $|R(\vec{\tilde{\alpha}}) - R(\vec\alpha)|$. 

\begin{claim}\label{obs-2}
$|R(\vec\alpha)|\geq \frac{1}{2^{(ndL)^{c_3}}}$ for some constant $c_3$.  
\end{claim}
 
\begin{proof}
Define the polynomial $\hat{R}(x_n) = R(\alpha_1, \ldots, \alpha_{n-1}, x_n ) = c \cdot \prod_{j=1}^{d'} (x_n-\beta_j)$ where $c$ is some constant and $d'\leq d$. Note that $\alpha_n$ is not a zero for $\hat{R}(x_n)$.  Consider the polynomial $Q(x_n) = p_n(x_n) \hat{R}(x_n)$. The set $\{\alpha_n, \beta_1, \ldots, \beta_{d'}\} \subseteq Z(Q)$ and $\alpha_n \neq \beta_j : 1\leq j\leq d'$. 
Using the root separation bound for $|\alpha_n-\beta_j|$ obtained in Lemma \ref{lemma-2}, we can easily lower bound that $|\hat{R}(\alpha_n)| \geq \frac{1}{2^{(n d L)^{c_3}}}$. 
\end{proof}

\begin{claim}\label{obs-3}
$|R(\vec{\tilde{\alpha}}) - R(\vec\alpha)|\leq 2^{(n d L)^{c_4}}$ for some constant $c_4$. 
\end{claim}

\begin{proof}
Define $R^{0}(\vec{\tilde{\alpha}})=R(\vec\alpha)$ and $R^{i}(\vec{\tilde{\alpha}}) = R(\tilde{\alpha}_1, \ldots, \tilde{\alpha}_i, \alpha_{i+1}, \ldots, \alpha_n)$. Then we use triangle inequality to notice that $|R(\vec\alpha) - R(\vec{\tilde{\alpha}})|\leq \sum_{i=1}^n |R^{i-1}(\vec{\tilde{\alpha}}) - R^i(\vec{\tilde{\alpha}})|$. Write explicitly 
$R^{i-1}(\vec{\tilde{\alpha}}) - R^{i}(\vec{\tilde{\alpha}}) = \sum_{\vec{e}} c_{\vec{e}} \tilde{\alpha}_1^{e_1}\ldots \tilde{\alpha}_{i-1}^{e_{i-1}} (\alpha_i^{e_i} - \tilde{\alpha}_i^{e_i}) \alpha_{i}^{e_{i+1}}\ldots\alpha_n^{e_n}$. 
Notice the upper bounds on $|\alpha_i| \leq 2^{(ndL)^{O(1)}}$, and $|\alpha_i - \tilde{\alpha}_i|\leq \epsilon$. We apply these bounds and use triangle inequality to get that 
$|R(\vec{\tilde{\alpha}}) - R(\vec\alpha)| \leq \epsilon \cdot 2^{(n d L)^{c_4}}$.
\end{proof}  
 
Combining Claim \ref{obs-2}, and Claim \ref{obs-3}, we get the lower bound 
$|f(\vec{\tilde{\alpha}})| \geq \frac{1}{2^{(n d L)^{c_3}}} - \epsilon \cdot (2^{(n d L)^{c_4}} + 2^{(n d L)^{c_2}})$. To make the calculation precise, let $3M = \frac{1}{2^{(n d L)^{c_3}}}$ and choose 
$\epsilon$ such that $\epsilon \cdot (2^{(n d L)^{c_4}} + 2^{(n d L)^{c_2}}) \leq M$. 

The final implication will be $|f(\vec{\tilde{\alpha}})|\leq M$ when $f\in I$ and $|f(\vec{\tilde{\alpha}})|\geq 2M$ when $f\not\in I$. It is important to note that the parameter $M$ can be pre-computed from the input parameters efficiently.  
 

Now we show how to verify that the guessed point $\vec{\tilde{\alpha}}$ is a good approximation of the roots for the univariate polynomials. We need to only verify that for each $i$, $\tilde{\alpha_i}$ is a good approximation for \emph{some root} of the univariate polynomial $p_i(x_i)$. The fact that it is also a good approximation for the non-zero of $R$ is already verified above. 
The $\NP$ machine, given $p_1, \ldots ,p_n$ guesses $\tilde{\alpha_i}$ using $b$ bits and verifies that $|p_i(\tilde{\alpha_i})| < 2^{-L} \epsilon^d $ which, by lemma \ref{npguess}, shows that the guessed $\tilde{\alpha_i}$ is $\epsilon$-close to some root of $p_i$. 

We note that such a guess always exists. 
Indeed, invoking Observation ~\ref{obs-1} with $|\alpha_i-\tilde{\alpha}_i|\leq \delta$ we can conclude that
$|p_i(\tilde{\alpha}_i)| \leq \delta \cdot 2^{(dL)^{{O}(1)}}$.
Now, the $\NP$ machine can guess $b$ bits such that $|\alpha_i-\tilde{\alpha}_i| \leq  2^{-b}$.
 We require $2^{-b} \cdot 2^{(dL)^{{O}(1)}} < 2^{-L} \epsilon^d$,
simplifying we get,   
$2^{-b} < 2^{-(dL)^{{O}(1)}} \cdot \epsilon^d$. Hence $b > (dL)^{{O}(1)} \log \frac{1}{\epsilon}$.
Thus using $\poly(d,L, \log \frac{1}{\epsilon})$ bits there is always a guess $\tilde{\alpha_i}$ for which $|p_i(\tilde{\alpha_i})| < 2^{-L} \epsilon^d$.  

\end{proof}
\end{document}